\documentclass[acus]{JAC2003}

\usepackage{graphicx}
\usepackage{booktabs}

\begin{document}
\title{TEST OF A 1.8 TESLA, 400 HZ DIPOLE FOR A MUON SYNCHROTRON\thanks{Work supported by 
NSF 757938 and  DOE  DE-FG05-91ER40622}}

\author{D. J. Summers\thanks{summers@phy.olemiss.edu}, L. M. Cremaldi, T. L. Hart, L. P. Perera, M. Reep, \\\ University of  Mississippi-\,Oxford, University, MS 38677, USA \\\
H. Witte, Brookhaven National Lab, Upton, NY 11973, USA \\\  S. Hansen, M. L. Lopes, Fermilab, Batavia, IL 60510, USA \\\  J. Reidy, Jr., Oxford High School, Oxford, MS 38655, USA}
\maketitle

\begin{abstract}
A 1.8\,T dipole magnet using thin grain oriented silicon steel laminations has 
been constructed as a prototype for a muon synchrotron ramping at 400\,Hz. Following the practice in 
large 3 phase transformers and our own Opera-2d simulations, joints are mitred to take advantage of 
the magnetic properties of the steel which are much better in the direction in which the steel was rolled. 
Measurements with a Hysteresigraph 5500 and Epstein frame show a high magnetic permeability which minimizes 
stored energy in the yoke allowing the magnet to ramp quickly with modest voltage. 
Coercivity is low which minimizes hysteresis losses.
A power supply with a fast
Insulated Gate Bipolar Transistor
(IGBT) switch and a capacitor was constructed. Coils are wound with 12 gauge copper wire.
Thin wire and laminations minimize eddy current losses.
The magnetic field was measured with a peak sensing Hall probe.
\end{abstract}

\section{Introduction}

A muon collider\,\cite{Neuffer} is perhaps unparalleled for exploring the energy frontier, if
an  economical design for muon cooling and acceleration can be finalized.   
Historically synchrotrons have provided low cost acceleration.
Here we present results on a dipole magnet prototype  for a relatively fast 400\,Hz synchrotron\,\cite{Summers}
for muons, which live for 2.2\,$\mu$s.  Low emittance muon bunches allow small apertures and
permit magnets to ramp with a few thousand volts, if the $B^{\,2}\!/2\mu$ energy stored in the magnetic yoke is kept low. 

\section{DIPOLE CONSTRUCTION AND RESULTS}

To minimize energy stored in the magnetic yoke, grain oriented silicon steel was chosen due to its high permeability as noted in Table 1.
Thin 0.011" 
AK Steel TRAN-COR H-1
laminations and 12 gauge copper wire minimize eddy current losses which 
go as
the square of thickness\,\cite{Sasaki}.
The copper wire will eventually be cooled with water flowing in stainless steel tubes\,\cite{West}.
The very low coercivity of grain oriented silicon steel as noted in Table 2 minimizes hysteresis losses\,\cite{Dawes}.
The power supply is an LC circuit with a 52\,$\mu$F polypropylene capacitor and a fast  IGBT 
Powerex CM600HX-24A switch. The magnet gap is 1.5 x 36 x 46 mm and 
$f = {1}/{2\pi \sqrt{LC}}$.  The energy stored in the gap is:
$$W = \int\!\!\frac{B^2}{2\mu_0} dh\,dw\,d\ell =\! \frac{LI^2}{2} =\! \frac{CV^2}{2} = 3.2\,{\rm{J}}$$  
An ideal dipole with N = 40 turns of copper wire and a current of 
$I = {B\,h}/{\mu_0N}$ = 54\,A would require a voltage of $V = 2 \pi B f N w \, \ell$ =\,315\,V
to generate 1.8\,T at 400\,Hz.

\vspace{-4mm}
\begin{table}[h!]
\centering
\caption{Relative permeability $(\mu/\mu_0)$ for 3\% grain oriented silicon steel\,\cite{Shirkoohi} as a function of angle to the rolling direction.
Dipole magnetic flux needs to be  parallel to the rolling direction.
The minimum at 1.3\,T and 55$^{\rm{o}}$ comes from the long diagonal (111) of the steel crystal.}  
\vspace{0.5mm}
\renewcommand{\arraystretch}{1.05}
\tabcolsep= 0.9mm
{\small
\begin{tabular}{@{}ccccccccc@{}} \hline
             & 0.1\,T     &  0.7\,T    & 1.3\,T & 1.5\,T & 1.7\,T & 1.8\,T & 1.9\,T & 2.0\,T\\ \hline
 $0^{\rm{o}}$   &29000&46000&49000&48000&30000&14000& 6000  & 180\\
 $10^{\rm{o}}$   &8000&14000&14000&10000&3000&&&\\
 $20^{\rm{o}}$  &3500&9800&9000&2100&&&&\\
 $55^{\rm{o}}$  &700&3800&540&&&&&\\
 $90^{\rm{o}}$  &660&3300  &2300& 320 & 120 & 80 & 60 & 50 \\ \hline
\end{tabular}}
\end{table}
\vspace{-4mm}
\begin{table}[h!]
\centering
\caption{Soft magnetic steel properties.}
\vspace{0.4mm}
\tabcolsep=0.6mm
\begin{tabular}{lccc} \hline
  Steel                                            &  $\rho$ &  H$_{\rm{c}}$ &  $\mu / \mu_0$ \\ 
                                              &  $\mu\Omega\cdot$cm &  \small Oersteds &   @B(T) \\ \hline 
Oriented 3\% silicon\,\cite{Shirkoohi}\rule{0pt}{10pt}                    &  46     & 0.09 & 14000@1.8\\
Ultra low carbon\,\cite{Laeger}     &  10     &  0.5    & 1900@1.5  \\
3\% silicon                 &  46     &   0.7  & 1600@1.3 \\
JFE 6.5\% silicon            &  82     &   0.2  & 1500@1.3 \\
 Hiperco 50A  ({\footnotesize{49\,Fe\,:\,49\,Co}})                                    &  42     & 0.3 & 2100@2.1\\ \hline
\end{tabular}
\vspace{-3mm}
\end{table}

\begin{figure}[h!]
   \centering
   \includegraphics*[width=60mm]{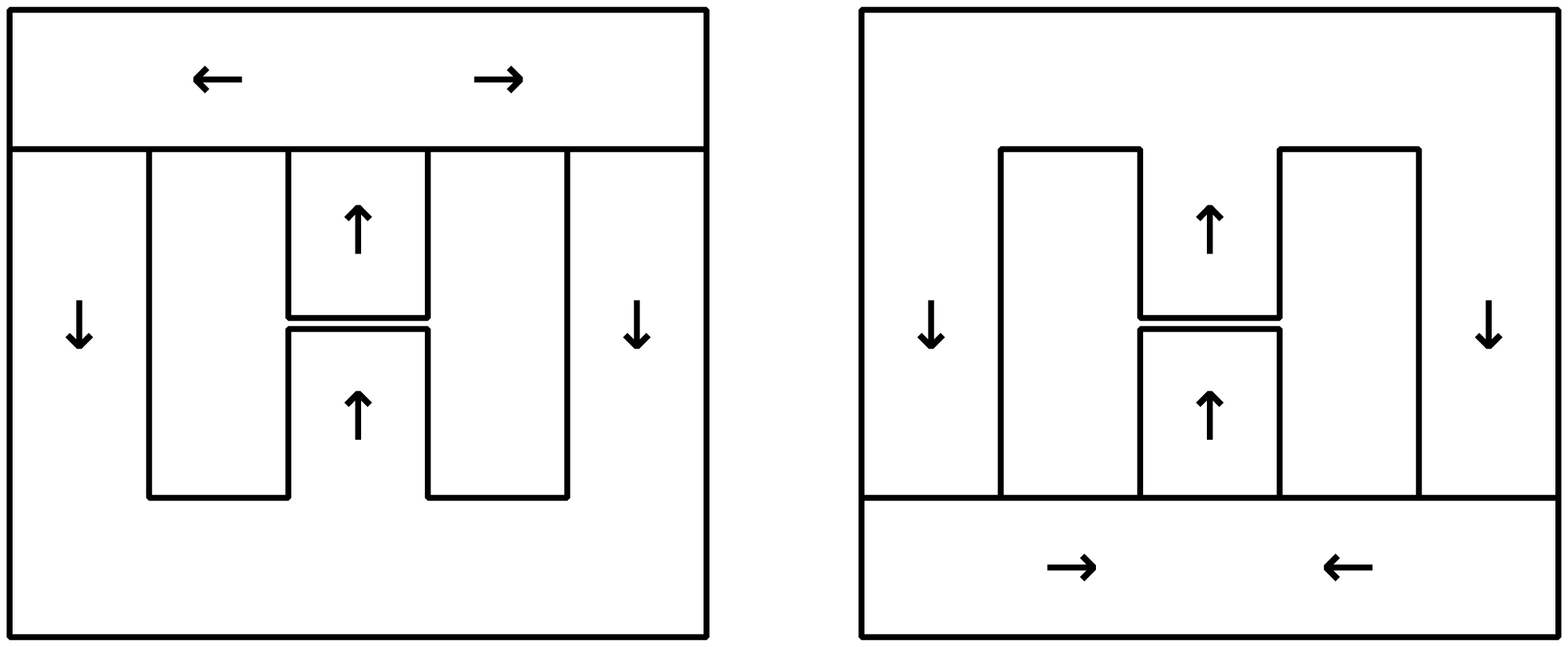}
   \includegraphics*[width=60mm]{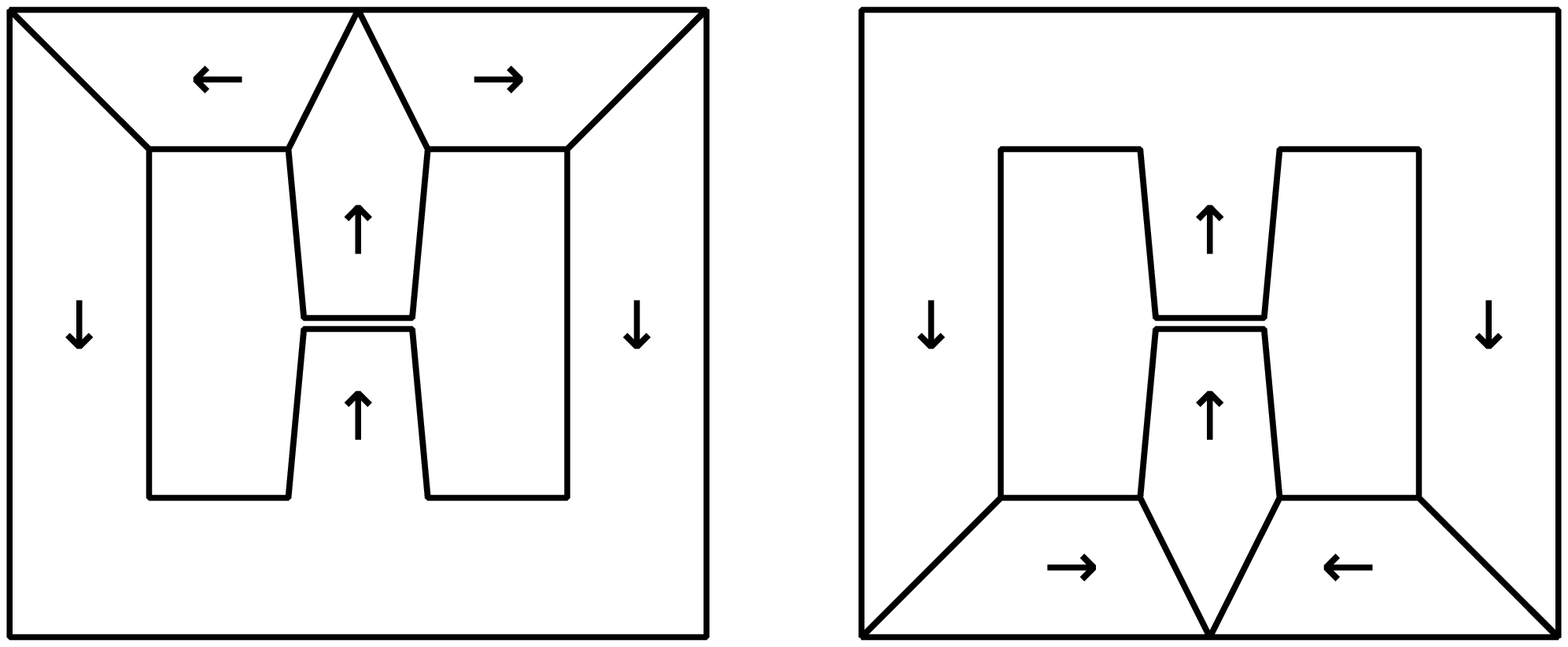}
   \vspace{-5mm}
   \caption{Dipole laminations with butt  and mitred joints.}
\end{figure}

Our first dipole prototype was made with butt joints as shown on the top of Fig.\,1.  The mitred joints in the second prototype
as shown on the bottom  of Fig.~1 work better. 

To further explore mitred joints, an Opera-2d simulation as shown in Fig.\,2 was run. 
Opera-2d only allows magnetic properties in the $x$ and $y$ directions to be entered. 
It then uses a  $\mu_\theta =  [(\cos{\theta}/\mu_{\,0^{\rm{\,o}}})^2  +  (\sin{\theta}/\mu_{\,90^{\rm{\,o}}})^2]^{-0.5}$ elliptical approximation.
This is problematic for  grain oriented silicon steel which has a minimum permeability at  $\mu_{\,55^{\rm{o}}}$.

\begin{figure}[h!]
   \centering
   \includegraphics*[width=76mm]{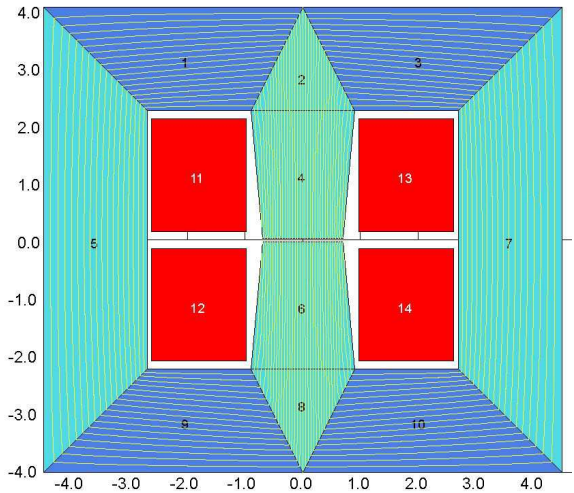}
   \vspace{-4mm}
   \caption{Opera-2d simulation of a mitred joint dipole. It worked, but there were some problems with convergence.}
\end{figure}

In two dimensions\,\cite{Fischer}:

$$ \frac{\partial^2 A}{\partial \, x^2} {\small{+}}  \frac{\partial^2 A}{\partial \, y^2}  \,- \frac{1}{\mu}\frac{\partial \mu}{\partial \, x} 
\frac{\partial A}{\partial \, x} \,- \frac{1}{\mu}\frac{\partial \mu}{\partial \, y} \frac{\partial A}{\partial \, y} = \,- \mu J_z$$
where
$\nabla\!\! \cdot\!\! A\! =\! 0$ and $A\!=\! A_z$. 
Thus    $B_x\! = - \frac{\partial A}{\partial y}$,   $B_y\! = - \frac{\partial A}{\partial x}$, and 
\ $\theta \!=\! {\rm{atan}}({B_y}/{B_x})$ are available for finite element iterations. The following subroutine generates a BH curve at any angle
using linear interpolation of a table with 5 angles.

{\scriptsize
\begin{verbatim}
      SUBROUTINE BH(NP,ANG,FANG,PM1,PM2,PM3,PM4,PM5,PM)
      IMPLICIT NONE
      INTEGER NP, J
      REAL FANG(5), PM(10), ANG, DANG
      REAL PM1(10), PM2(10), PM3(10), PM4(10), PM5(10)   
C       
      IF(ANG.GE.FANG(1) .AND. ANG.LT.FANG(2)) THEN
      DO 10 J=1,NP
      DANG = (ANG - FANG(1))/(FANG(2) - FANG(1))
  10  PM(J) = PM1(J) + DANG*(PM2(J) - PM1(J))
      ELSE IF(ANG.GE.FANG(2) .AND. ANG.LT.FANG(3)) THEN
      DO 20 J=1,NP
      DANG = (ANG - FANG(2))/(FANG(3) - FANG(2))
  20  PM(J) = PM2(J) + DANG*(PM3(J) - PM2(J))
      ELSE IF(ANG.GE.FANG(3) .AND. ANG.LT.FANG(4)) THEN
      DO 30 J=1,NP
      DANG = (ANG - FANG(3))/(FANG(4) - FANG(3))
  30  PERM(J) = PM3(J) + DANG*(PM4(J) - PM3(J))
      ELSE IF(ANG.GE.FANG(4) .AND. ANG.LE.FANG(5)) THEN
      DO 40 J=1,NP
      DANG = (ANG - FANG(4))/(FANG(5) - FANG(4))
  40  PM(J) = PM4(J) + DANG*(PM5(J) - PM4(J))
      END IF
      RETURN
      END
\end{verbatim}}

To check our steel, the hysteresis loop shown in Fig.\,3 was measured with  an Epstein frame
and Hysteresigraph 5500.
Fig.\,4 shows our magnet with mitred joints and Fig.\,5 shows a  permanent magnet used to check Hall probes.
DC magnetic tests were run on our two magnets as shown in Fig.\,6. The mitred joint magnet starts to become nonlinear at 1.7\,T.
Fig.\,7 shows our fast ramping IGBT power supply. Figs.\,8 and 9 show results of ringing our mitred joint dipole.

\begin{figure}[h]
   \centering
   \includegraphics*[width=82mm]{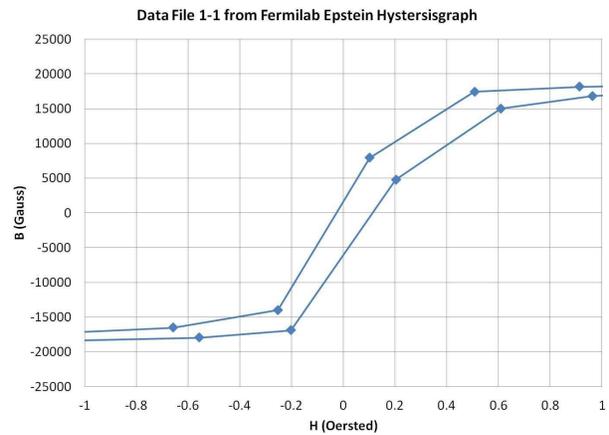}
   \vspace{-8mm}
   \caption{BH(0$^{\rm{o}}$) for  our steel. $B^{\,2}\!/2\mu$ and $H_c$  are  small.  $\mu / \mu_0$~=~17000 and 9000 at  1.7\,T and 1.81\,T, respectively.}
\end{figure}

\begin{figure}[h!]
   \centering
   \includegraphics*[width=82mm]{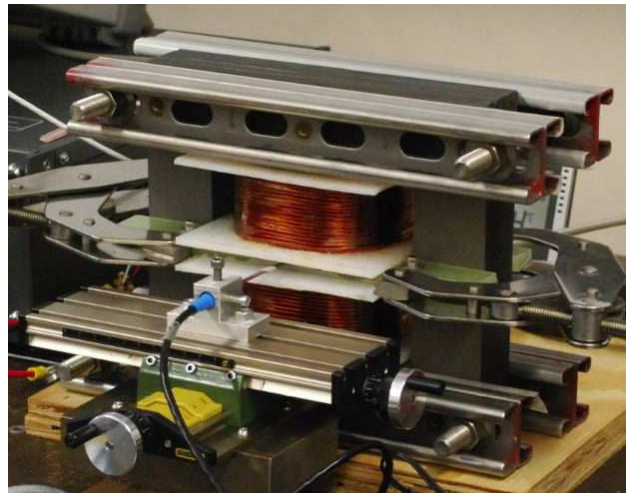}
   \caption{Dipole magnet with mitred  laser cut laminations. Laminations were reannealed and recoated after cutting.}
\end{figure}

\begin{figure}[h!]
   \centering
   \includegraphics*[width=60mm]{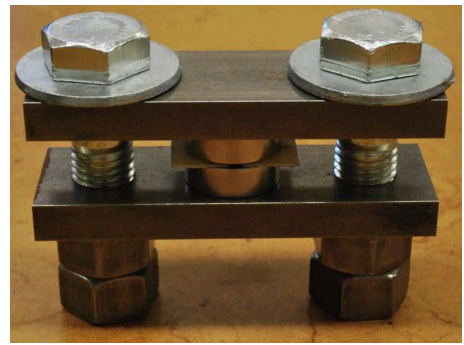}
   \caption{1.17\,T \, NdFeB magnet with 38\,mm diameter pole faces used to check Hall probes.}
\end{figure}

\begin{figure}[h]
   \centering
   \includegraphics*[width=80mm]{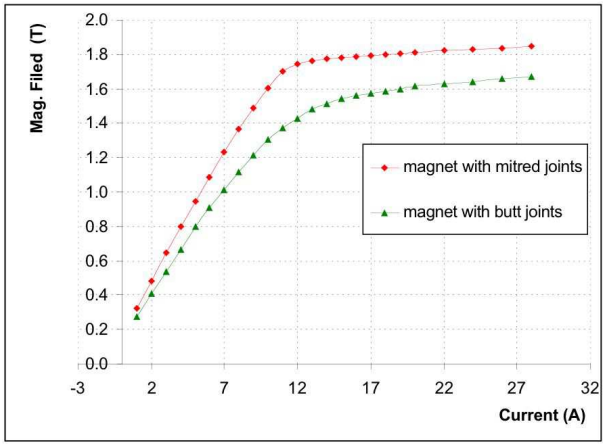}
   \caption{B field vs.~current for our butt joint magnet (bottom)  and mitred joint magnet\,(top).  
   B\&K Precision 1794 DC linear power supply and F.\,W.\,Bell 5180 Hall probe.}   
\end{figure}

\begin{figure}[h]
   \centering
   \includegraphics*[height=45mm,angle=90]{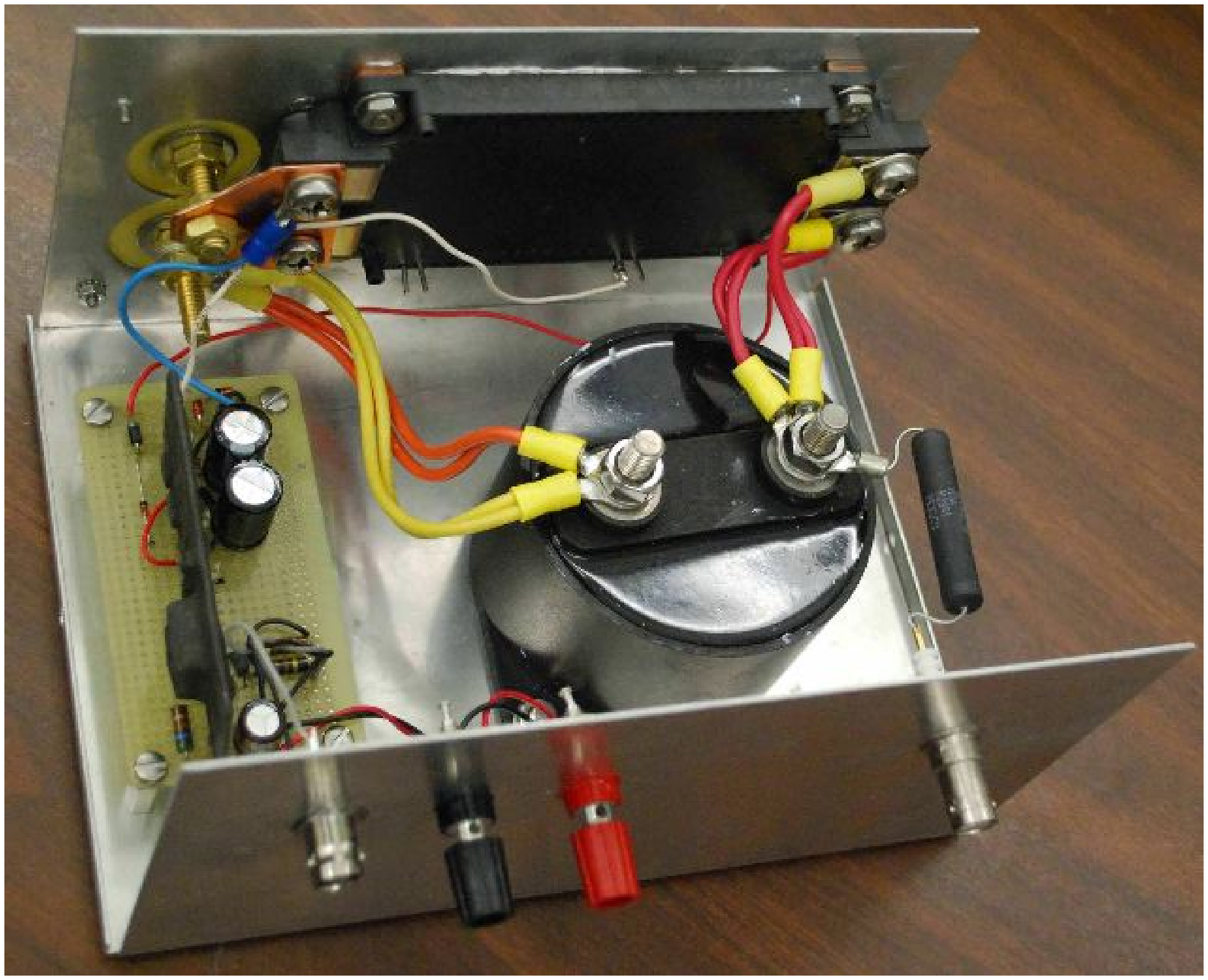}  
    \includegraphics*[height=56mm]{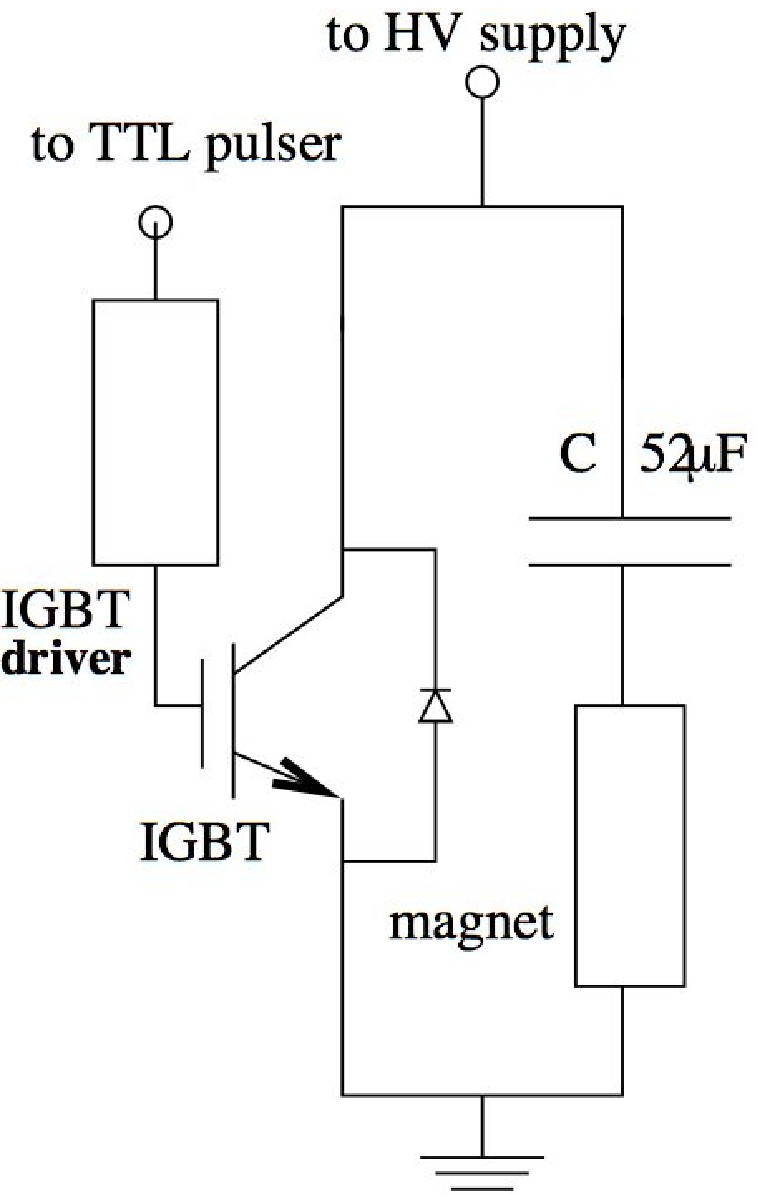}      
   \vspace*{-2mm}
   \caption{Fast  IGBT\,/\,52\,$\mu$F capacitor power supply. Powerex CM600HX-24A IGBT and  VLA500\,-\,01 gate driver. }
\end{figure}

\begin{figure}[h]
   \centering
   \includegraphics*[width=82mm]{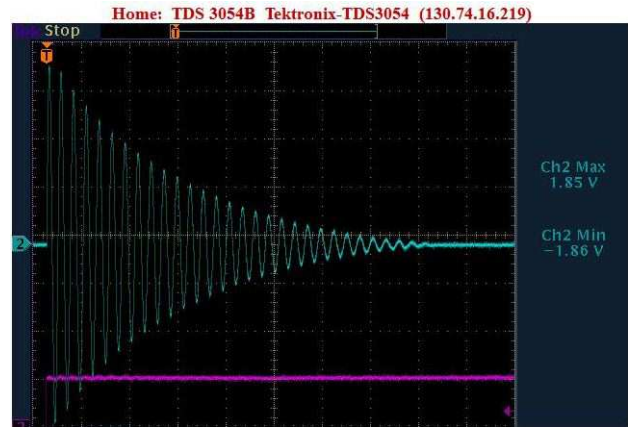}
   \caption{Mitred dipole with 53 turns and 430\,V gives 1.81\,T at   425\,Hz.  Energy loss is 10\% per half cycle. 
   F.\,W. Bell 5180 peak sensing Hall probe connected to a Tektronics TDS3054B 
oscilloscope.}    
    \end{figure}

\begin{figure}[h]
   \centering
   \includegraphics*[width=82mm]{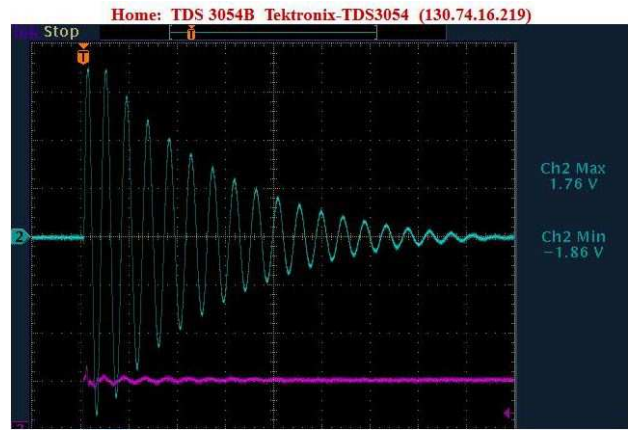}
      \caption{Mitred dipole with 18 turns and 550\,V gives 1.81\,T at   1410\,Hz.  Energy loss is 15\% per half cycle. }   
\end{figure}

%http://mafurman.lbl.gov/LBNL-38563.pdf

\section{SUMMARY}

A 1.8\,T dipole can run at 400\,Hz. A magnetic flux circuit with a large yoke path to gap ratio works with high permeability steel. 
The next step is improving field quality and the accuracy of pole faces,
as well as matching calculated and observed losses.
Simulation of anisotropic steel has proven challenging. Transverse beam pipe impedance, which is proportional  
to the inverse cube of beam pipe diameter, will probably dictate a 12\,mm dipole gap.
Radiation damage of steel needs to be explored\,\cite{Park} and a Rogowski profile needs to be added to magnet ends.

\section{ACKNOWLEDGMENT}
Finally, we are most grateful to  S.\,Berg, 
K.\,Bourkland,
A.\,Garren, 
K.\,Y.\,Ng,
R.\,Palmer,
R.\,Riley,
A.\,Tollestrup,
J.\,Tompkins, 
S.\,Watkins,
and J.\,Zweibohmer
for their help.

\end{document}